\documentclass{an}
\usepackage{graphicx}
\usepackage{times}
\usepackage{xspace}
\Volume{00}              
\Year{0000}              
\Month{00}               
\Pagespan{000}{000}      

\newcommand{\arcdeg}{\mbox{$^\circ$}\xspace}
\renewcommand{\arcmin}{\mbox{$^\prime$}\xspace}

\newcommand{\Ms}{\mbox{$M_{\odot}$}\xspace}

\def\la{\hbox{\rlap{\raise.3ex\hbox{$<$}}\lower.8ex\hbox{$\sim$}\ }}
\def\ga{\hbox{\rlap{\raise.3ex\hbox{$>$}}\lower.8ex\hbox{$\sim$}\ }}
\def\deg{$^{\circ}$~}

\begin{document}
\lhead[\thepage]{Grindlay,J.~et~al: Chandra Multi-wavelength Plane
(ChaMPlane) Survey:  Design and Initial Results}
\rhead[Astron. Nachr./AN~{\bf XXX} (200X) X]{\thepage}
\headnote{Astron. Nachr./AN {\bf 32X} (200X) X, XXX--XXX}

\title{Chandra Multi-wavelength Plane (ChaMPlane) Survey: Design and
Initial Results}

\author{Jonathan Grindlay$^1$, Ping Zhao$^1$, JaeSub Hong$^1$, 
Johnathan Jenkins$^1$, Dong-Woo Kim$^1$, Eric Schlegel$^1$, Jeremy Drake$^1$,
Vinay Kashyap$^1$, Peter Edmonds$^1$, Haldan Cohn$^2$, Phyllis Lugger$^2$ and
Adrienne Cool$^3$}
\institute{
$^1$Harvard-Smithsonian Center for Astrophysics, 60 Garden St.,
Cambridge, MA 02138 \\
$^2$Department of Astronomy, Indiana University, Swain Hall West,
Bloomington, IN 47405 \\
$^3$Department of Physics and Astronomy, San Francisco State University,
San Francisco, CA 94132
}
\date{Received {date will be inserted by the editor}; 
accepted {date will be inserted by the editor}} 
\abstract{
The Chandra Multiwavength Plane (ChaMPlane) Survey of the
galactic plane incorporates serendipitous sources from selected
Chandra pointings in or near the galactic plane 
($b$$<$12\arcdeg; $\ge$20 ksec; lack
of bright diffuse or point sources) to measure or constrain the
luminosity function of low-luminosity accretion sources in the Galaxy.
The primary goal is to detect and identify accreting white dwarfs
(cataclysmic variables, with space density still uncertain by a factor
of $>$10-100), neutron stars and black holes (quiescent low mass X-ray 
binaries) to constrain their space densities and thus origin and
evolution. Secondary objectives are to identify Be stars in high mass
X-ray binaries and constrain their space densities, and to survey the
H-R diagram for stellar coronal sources. A parallel optical imaging 
under the NOAO Long Term Survey program provides deep optical images
using the Mosaic imager on the CTIO and KPNO 4-m telescopes. The 
36\arcmin$\times$36\arcmin optical images (H$\alpha$, R, V and I) 
cover $\sim$ 5$\times$ the area of each
enclosed Chandra ACIS FOV,  providing an extended survey of emission
line objects for comparison with Chandra. Spectroscopic followup of
optical counterparts is then conducted, thus far with WIYN and
Magellan. The X-ray preliminary results from both the Chandra and
optical surveys will be presented, including logN-logS vs. galactic
position (l,b) and optical idenifications. 
}
\correspondence{josh@cfa.harvard.edu}

\maketitle

\section{Introduction and Objectives}

The remarkable angular resolution of the Chandra X-ray 
Observatory enables the first high sensitivity survey of low 
luminosity accretion sources in the Galaxy for which optical 
identifications, and thus source content, can be reasonably 
expected. The arcsec source positions allow relatively 
unambiguous optical (and IR) identifications even in the 
most crowded galactic fields (not including dense clusters), 
and the point source sensitivity for sources within $\sim$4\arcmin 
of the field center allows sources with power law spectra 
(with photon index $\Gamma$ = 1.7,  as expected for both low- and 
high-mass X-ray binaries) to be detected at galactic bulge 
distances ($\sim$8 kpc) with ``hard" X-ray luminosities 
$L_X$(2--8keV) $\sim$ 2$\times$10$^{31}$~erg/s. This, in turn, 
enables the detection of 
the bright-half of the luminosity distribution for accreting 
white dwarfs (CVs), nearly the full distribution (as presently known) 
of qLMXBs containing either neutron stars (NSs) or black holes (BHs), 
and the full distribution (again, as known) of accreting high 
mass X-ray binaries (HMXBs) such as the accreting Be 
binary systems. This capability allows a major advance in 
previous galactic surveys, which most recently have been 
conducted with the ROSAT all sky survey (Voges et al 
1993) and the Einstein galactic plane survey (Hertz, Bailyn, 
Grindlay et al 1990).  XMM is of course also now able to 
contribute in a major way to the study of faint galactic X-ray 
source populations, and preliminary results have been 
reported by Motch et al (2002) and Warwick (these 
proceedings). 

       We have initiated the Chandra Multiwavelength Plane 
(ChaMPlane) survey to probe the luminosity functions and 
space density of cataclysmic variables (CVs) and quiescent 
low mass X-ray binaries (qLMXBs) containing either 
neutron stars (NSs) or black holes (BHs). Detecting CVs on 
galactic scales would constrain the origin (are they all from 
primordial binaries?) and evolution (which ones produce 
SNIa's?) of these most common accretion-powered sources.  
The qLMXBs, in turn, are the reservoir of even more exotic 
binary evolution processes leading, alternatively, to the 
millisecond pulsars and the largest known samples of stellar 
mass black holes.  Secondary objectives for ChaMPlane 
include measuring and mapping the other major reservoir of 
accreting binaries, the Be systems with NS (and possibly 
BH) companions; and the search for isolated BHs accreting 
from giant molecular clouds. Finally, ChaMPlane will 
acquire perhaps the largest sample of identified stellar 
coronal sources and enable measurements or limits of X-ray 
activity of stars across the H-R diagram.

\section{Survey Design and Expectations}

The X-ray survey is derived from serendipitous sources 
detected in  Chandra observations of galactic fields with 
nominal exposure times  $\ge$20--30ksec, 
galactic latitude $b$$<$12\arcdeg, and
lack of bright sources, or extended emission in the 
field which would reduce sensitivity. The observations are 
restricted to those taken with the ACIS-I or -S instruments 
on Chandra to allow for spectral coverage to harder energies 
(important for the often-absorbed galactic fields) and at least 
hardness ratio (X-ray colors) if not actual spectral analysis. In 
order to achieve significant coverage of regions of the 
galactic plane and to acquire significant samples, the  
original proposed goal was to obtain $\sim$100 Chandra fields. 
Given the rate at which ``clean" galactic fields are being 
observed with Chandra, this is expected to take 5 years to 
acquire. The X-ray data processing is done with the same 
XPIPE scripts as developed (Kim et al 2003) for the high 
latitude ChaMP survey. Thus far data have been processed 
for some 40 fields (of which 15 are reported here) in the 
same 3 bands as for ChaMP (cf. Fig. 3 below) 
for analysis of  X-ray fluxes and colors.

 A parallel deep optical imaging survey is being 
conducted with the CTIO and KPNO 4m telescopes and 
large field (36\arcmin) Mosaic cameras under the NOAO 
Long Term Surveys Program (see Zhao et al, these 
proceedings). This provides the required deep optical image 
for initial source identification (and followup spectroscopy) 
but is also designed to do an initial selection for likely 
accretion-powered objects by imaging in H$\alpha$ vs. R to look 
for ``blue" objects in the (H$\alpha$--R) vs. R color magnitude 
diagram, given that H$\alpha$ is a (nearly) ubiquitous feature of 
accretion disks or flows. To provide additional constraints 
on the possible counterpart's spectral type and reddening 
(particularly if followup spectra are not possible), additional 
images in V and I are acquired so that a (V-R) vs. (R-I) 
plane can be constructed. Unfortunately the reddening vector 
is relatively closely aligned with the main sequence in this 
color system so that photometric classifications are relatively 
coarse if relative or differential reddening is not known. The 
red filter system (V,R,I) is itself chosen to minimize 
extinction and to provide the comparison R band for H$\alpha$. 
Total exposure times in H$\alpha$, R, V and I are $\sim$2.5h, 0.5h, 0.3h 
and 0.3h, respectively, to reach $\sim$5\% photometry at R$\sim$24 
(10\% in H$\alpha$) which allows a threshold 
of EW(H$\alpha$) $\sim$15\AA\ for 
the 80\AA\ wide H$\alpha$ filter in the Mosaic cameras. Photometry 
has now been obtained on 24 fields, or $\sim$1/4 the desired total.

 The CV space density has been estimated recently to be 
$\sim$10$^{-5}$ pc$^{-3}$ (Patterson 1998) 
but this is still uncertain by an 
order of magnitude (either way) and is largely based on both 
optical and X-ray surveys typically within 1kpc. The galactic 
distribution is essentially unknown but could be probed by 
ChaMPlane, for the brighter CVs, out to the galactic bulge. 
For a likely CV scale height of $\sim$200pc (e.g. Schwope et al 
2002), the predicted total number interior to the Sun is $\sim$ 
4$\times$10$^5$. The local CV space density extrapolated to the galactic 
center would suggest a total ChaMPlane-CV ``column" of 
$\sim$20, of which perhaps $\sim$ 3--5 might be detected given $N_H$ and 
luminosity distributions. The likely much larger population 
of ``quiescent" CVs (with probable 
$L_X$ $\le$ 10$^{30}$ erg/s) suggested 
by Howell, Rappaport and Politano (1997) and Townsley 
and Bildsten (2002) could be detected to $\sim$1-3kpc and 
contribute a comparable total. 

 The qLMXB estimates are 
based on millisecond pulsar surveys and soft X-ray transients 
for the NS and BH systems, respectively, with uncertainties 
due to both source evolution and duty cycles. The total 
population interior to the solar circle is probably $\ge$100$\times$ 
below the CV numbers, yielding perhaps 0.1--1 per 
ChaMPlane field, but again the uncertainties are large. 
Similar totals, or perhaps larger, obtain for estimates of the 
Be binary populations expected. The above estimates 
suggest that ChaMPlane could double (at least) the total 
numbers of accretion-powered X-ray binaries when 
identifications are completed with our followup 
spectroscopy. Expected numbers of previously unrecognized 
objects, such as isolated $\sim$10\Ms BHs accreting from 
GMCs, are much more difficult to predict (Agol and 
Kamionkowski 2002) and will likely require IR photometry 
and spectroscopy. An important discriminant throughout 
between compact objects and coronal stars is the hard X-ray 
spectrum ($>$2--3 keV) expected from accreting compact 
objects but not (except in flaring) stars. 

\section{Initial X-ray Results}

In Figure 1 we show the galactic distribution of the first 
sample of 15 ChaMPlane fields analyzed.
\begin{figure}
\resizebox{\hsize}{!}
{\includegraphics[bb=120 98 487 710,clip=true,angle=-90]{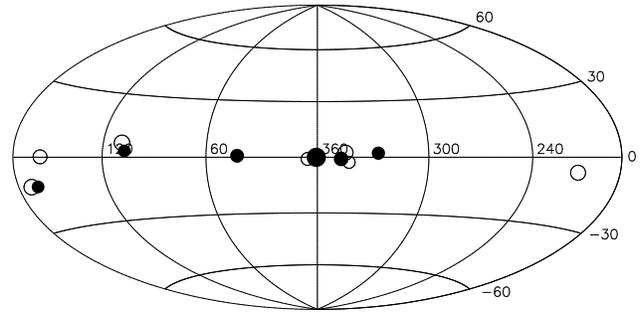}}
\caption{Initial ChaMPlane ACIS-I (solid) and -S (open) fields in 
galactic coordinates.}
\label{Fig1}
\end{figure}
We first derive logN-logS, the integral 
number-flux counts, for comparison with other galactic and high 
latitude surveys. 
For initial analysis, sources selected from within the central 4\arcmin 
of the Chandra field center (to minimize off-axis 
degradation of point spread function and sensitivity) and 
with at least 10cts in the broad band (0.3--8keV). 
For each field, we derive the corresponding flux for 
this detection limit and accumulate the effective sky areal 
coverage as a function of flux limit. Source counts detected 
are converted to flux assuming a power law spectrum ($\Gamma$ = 
1.7), appropriate for CVs, qLMXBs or AGN,  and are 
corrected for absorption by the nominal $N_H$ for that field. 
Results are shown in Figure 2 for two cuts on galactic 
longitude for the 9 fields with  $\ell \le 44\arcdeg$. \\ \\
\indent The excess sources in the galactic bulge are distributed with 
-1 slope appropriate to a disk distribution vs. the {-3/2} slope 
over the same flux range exhibited by the AGN counts from 
ChaMP (Green et al, these proceedings), plotted as the 
lower curve in both figures and dashed line 
(top figure), which connects the Chandra Deep 
Field North and brighter flux ASCA counts. The galactic 
source population is at least a factor of 3-5 over the AGN 
counts and is unlikely to be due to enhanced stars in the 
bulge since these would be expected to be predominantly 
soft sources (though pre-main sequence stars 
may contribute). The X-ray colors and limits plotted in Figure 3 show 
that most of these bulge sources are 
consistent with hard and absorbed spectra 
as expected for accretion sources. 
The two points at lower right may be sources with both 
soft and hard spectral components, such as found in 
magnetic CVs, and would be foreground sources. 
\begin{figure}
\resizebox{\hsize}{!}
{\includegraphics{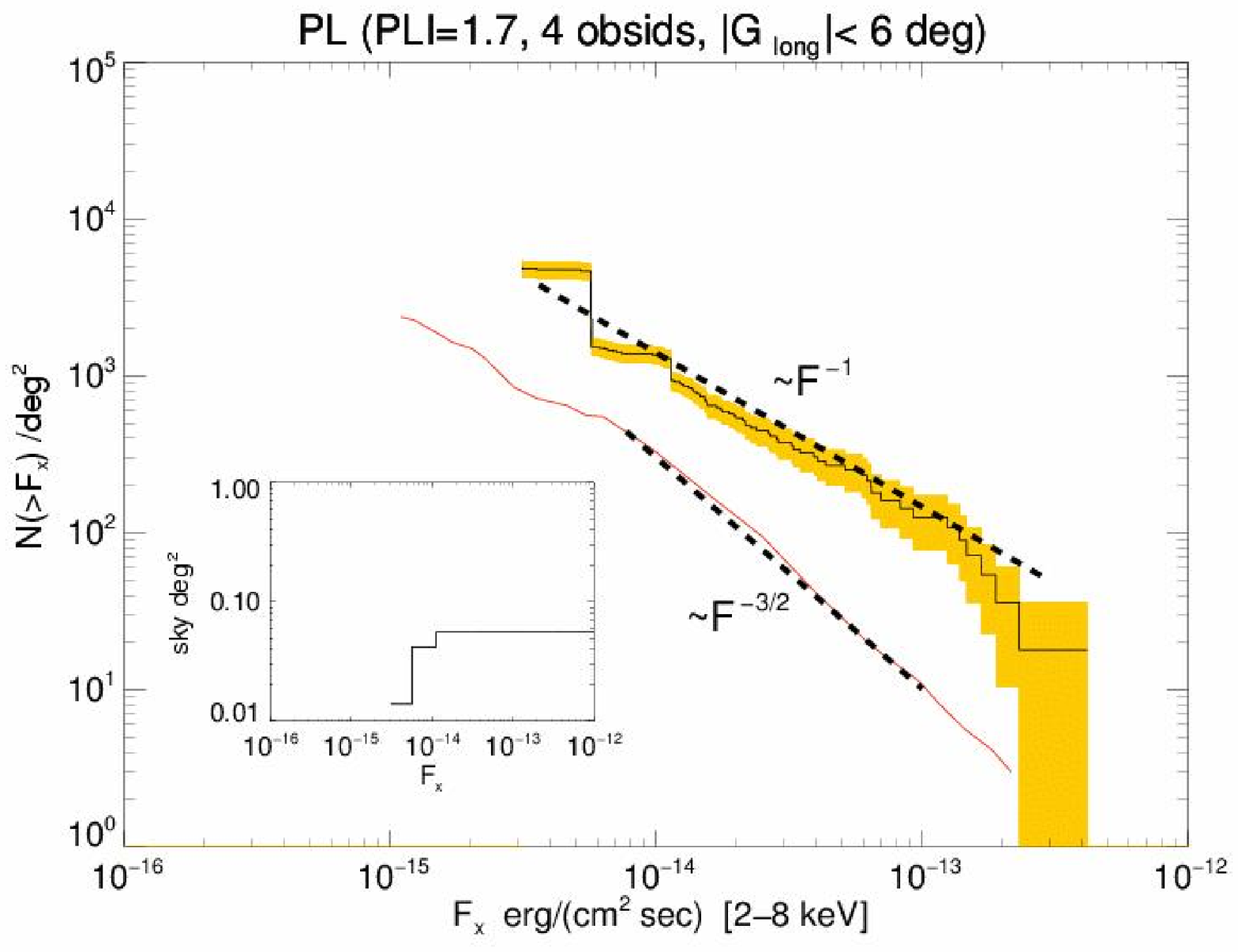}} \\
\resizebox{\hsize}{!}
{\includegraphics{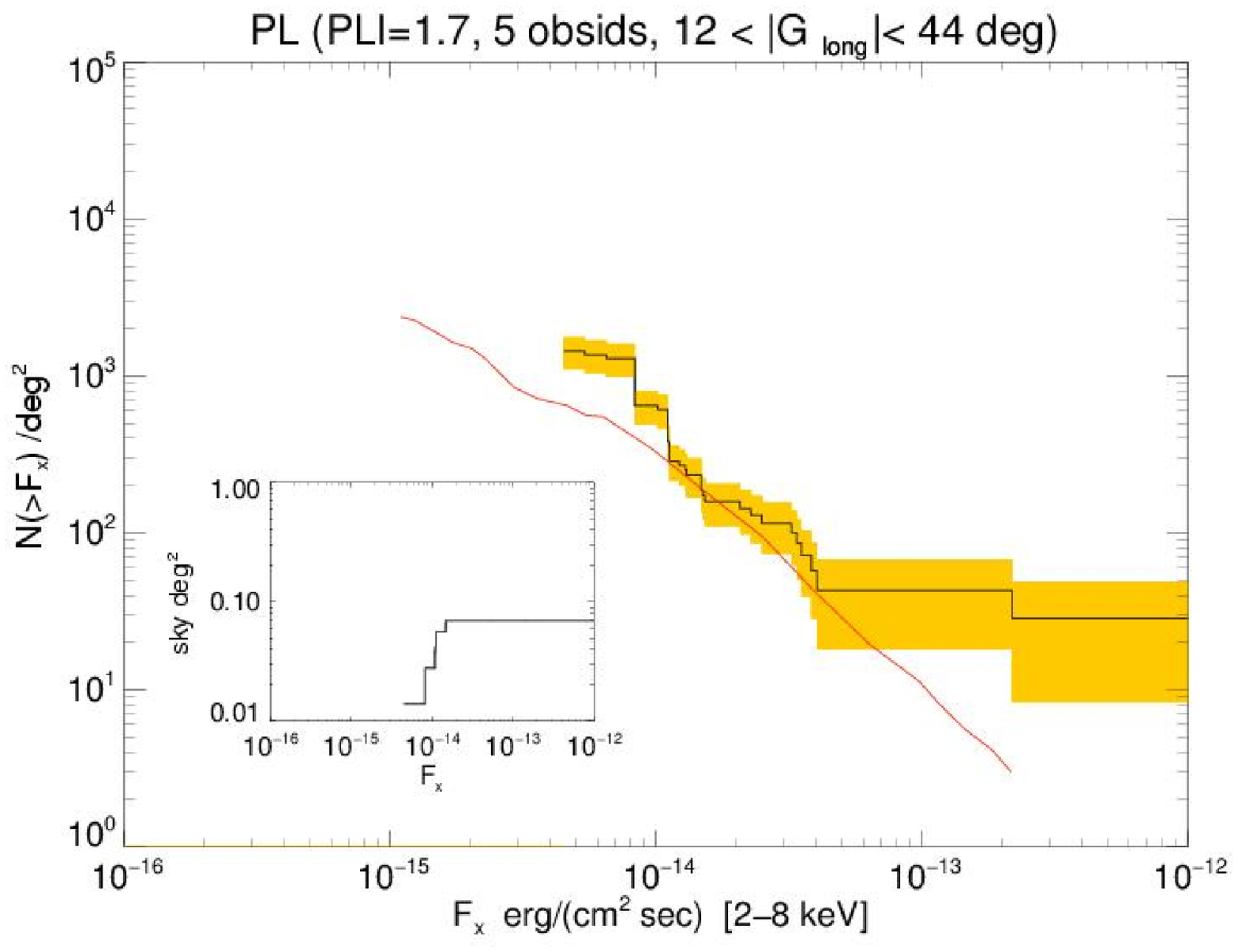}}
\caption{Preliminary ChaMPlane LogN-LogS and error band  
for sources in central (\la4\arcmin) Chandra fields in galactic 
longitude ranges shown compared with high latitude 
AGN counts.}
\label{Fig2}
\end{figure}
\begin{figure}
\resizebox{\hsize}{!}
{\includegraphics[angle=90]{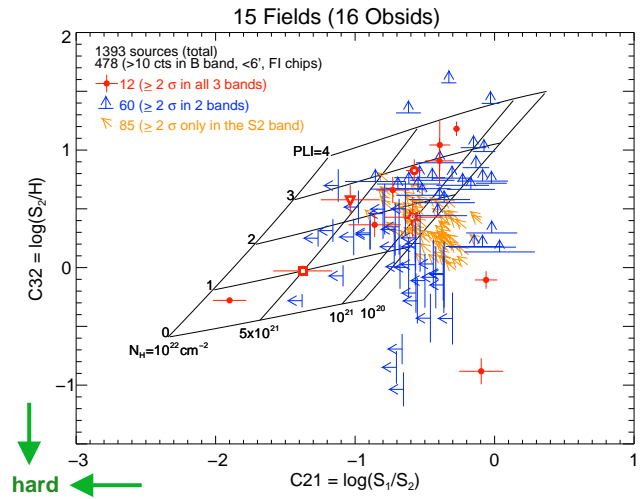}}
\caption{Source X-ray colors and  limits vs. PL models and $N_H$. Reduced 
source counts from 1393 total due to 
restriction $\le$6\arcmin (from center) and 
separate detections in individual bands. Bands are: S1=0.3--1.2keV, 
S2=1.2--2.5keV, H=2.5--8keV, B=0.3--8keV.}
\label{Fig3}
\end{figure}
\begin{figure}
\resizebox{\hsize}{!}
{\includegraphics{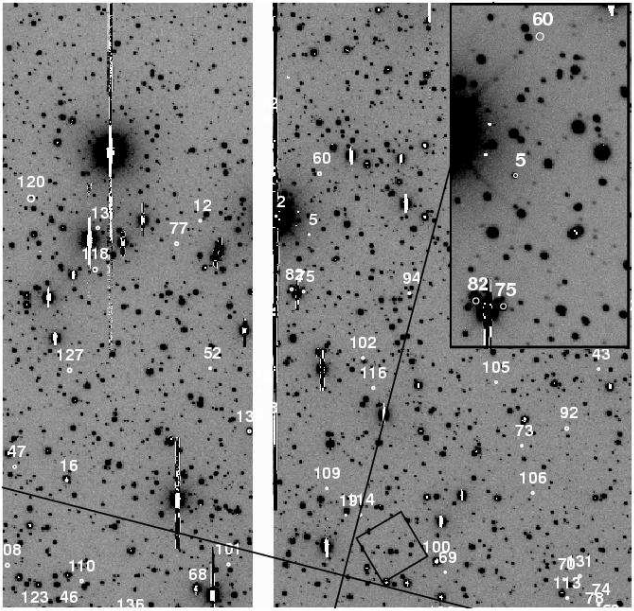}}
\caption{R image ($\sim$ 8\arcmin$\times$8\arcmin) 
vs. Chandra IDs for Galactic Center Arc 
field and new CV (Chandra \#5; cf. inset). N-right; E-up.}
\label{Fig4}
\end{figure}

\section{Initial Optical Results}

In Fig. 4 we show the partial R image for the Galactic Center 
Arc field (Mosaic; showing a chip gap) with ACIS-I chip 
overlay (black lines) and central pointing position (diamond) 
together with Chandra sources (numbers) and their 2$\sigma$ error 
circles. The image shows both that about half of the sources 
have optical IDs in this image (to R$\sim$23) for which the 
reddening is large (log($N_H$) $\sim$ 22.1). The  (H$\alpha$-R) vs. R color 
magnitude diagram for this field (Fig. 5) shows that two 
Chandra sources are significant H$\alpha$ objects. 
\begin{figure}
\resizebox{\hsize}{!}
{\includegraphics{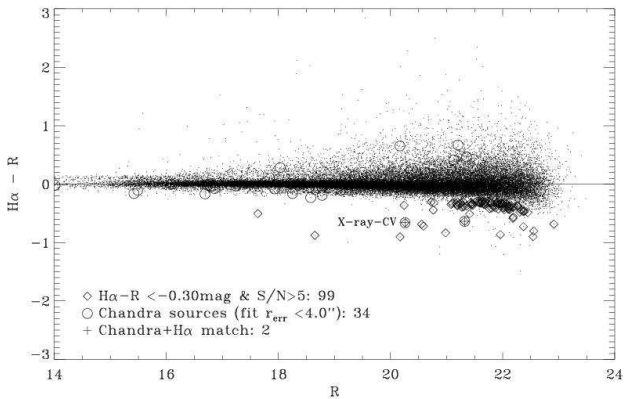}}
\caption{CMD identifying  H$\alpha$ objects 
with 2 Chandra sources; others with similar significance are mostly outside 
the Chandra FoV.}
\label{Fig5}
\end{figure}
 The R $\sim$ 20.2 object labelled X-ray CV is source \#5 in Fig. 
4 and enlarged in the inset in upper right corner:  
the 0.6$\,^{\prime\prime}$ radius circle shows the excellent 
positional astrometry, while \#60 shows a typical blank field (R $\ge$ 23.5)
source. Source 5 is 
confirmed as an emission line object and probable CV in our 
Magellan spectrum shown in Fig. 6.
\begin{figure}
\begin{center}
\resizebox{0.9\hsize}{!}
{\includegraphics[bb=54 87 569 675,angle=-90,clip=true]{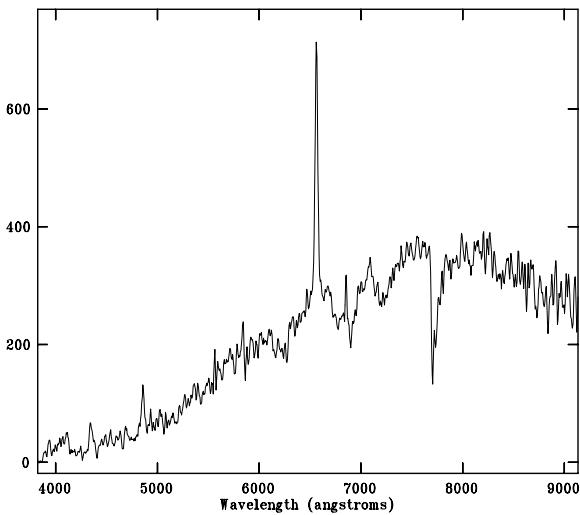}}
\end{center}
\caption{Spectrum of the first ChaMPlane X-ray selected CV.}
\label{Fig6}
\end{figure}
The spectrum shows Balmer line emission broadend beyond 
the instrumental resolution as well as weak HeI emission 
suggesting a reddened dwarf nova. Its X-ray/optical flux ratio 
is $F_X$/$F_R$ $\sim$ 0.2 and is plotted vs. R in Fig. 7 
as the square symbol along with values for the $\sim$~300 
sources identified in these 15 fields. In Fig. 3 it is 
consistent with a foreground CV with log($N_H$) $\sim$~21 and 
photon index 1 (or a brems spectrum).  
The point labelled Be? is a second spectroscopically 
confirmed counterpart: a probable Be HMXB system.   
Its relatively large value of $F_X$/$F_R$ $\sim$ 0.002 
suggests it is an accretion-powered source, probably 
containing a neutron star. The fact that the 
bulge sources have systematically lower $F_X$/$F_R$ values than 
the anti-center sources is likely due to differential reddening: 
both $F_R$ and $F_X$ values were ``de-absorbed'' by the full $N_H$ in each field. 
The plotted R~\ga10 magnitudes were actually observed as R~\ga14. 
\begin{figure}
\resizebox{\hsize}{!}
{\includegraphics{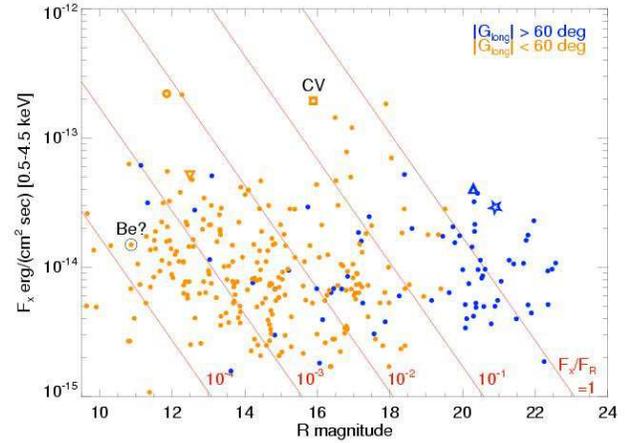}}
\caption{Un-absorbed X-ray/optical flux vs. R (de-reddened) 
for initial sample of 
ChaMPlane sources inside vs. outside $\ell$ = 60\deg.}
\label{Fig7}
\end{figure}


\section{Conclusions}

ChaMPlane has found dramatic evidence for a new population of low 
luminosity hard X-ray sources (HXs) in the central galactic bulge. 
Young stellar objects, found as HXs in SgrB2 by Takagi et al (2002),  
may not be numerous enough but models and stellar 
population studies (e.g. van Loon et al 2002) are needed. 
The source counts at 12\deg $\le \ell \le$ 44\deg  show a flattened 
disk component at $F_X$ \ga 10$^{-13}$~cgs that is 
consistent with the point source population at $\ell \sim$ 21\deg 
found by Motch et al (2002). However this ``bright'' galactic disk 
distribution is exceeded by the steeper AGN counts at fluxes  
$F_X$ \la 10$^{-13.5}$~cgs, only to be exceeded by the 
still larger normalization disk population in the central bulge. 
This may resolve the discrepancy between the claims for AGN 
dominating the faint ($\sim3\times10^{-15}$ cgs) 
galactic source distribution at $\ell$ = 28.5\deg (Ebisawa et al 2001)
vs. an excess in the lower-sensitivity wide-field 
galactic center survey (Wang et al 2002). 
ChaMPlane  will map out this distribution, and 
our upcoming deep pointing on Baades Window with 
Chandra and HST may reveal the nature of this bulge population.  

\acknowledgements

We thank  ChaMP colleagues 
for assistance, D. Hoard, S. Wachter and T. Abbott for 
help at CTIO,  and the NOAO Surveys and 
Magellan  staff for support. We acknowledge 
Chandra  grants AR1-2001X, AR2-3002A and 
NSF grant AST-0098683.


\end{document}